\documentclass{iaus}
\usepackage{graphicx}

\title[Inward bound: following gas\ldots] 
{Inward bound: following gas flows from nuclear spirals to the accretion disk}

\author[T.~Storchi-Bergmann]   
{Thaisa Storchi-Bergmann}%

\affiliation{Instituto de F\'{\i}sica, Universidade Federal do Rio Grande do Sul, Porto Alegre, RS, Brazil \\[\affilskip] email: thaisa@ufrgs.br}

\pubyear{2006}
\volume{238}  
\pagerange{001--999}
\date{??? and in revised form ???}
\jname{Black Holes: from Stars to Galaxies -- across the Range of Masses}
\editors{V. Karas \& G. Matt, eds.}
\begin{document}

\maketitle

\begin{abstract}
A recent analysis of HST optical images of 
34 nearby early-type active galaxies 
and of a matched sample of 34 inactive galaxies
-- both drawn from the Palomar survey -- shows a clear excess of nuclear 
dusty structures (filaments, spirals and disks) 
in the active galaxies. This result supports the
association of the dusty structures with the material which
feeds the supermassive black hole (hereafter SMBH). 
Among the inactive galaxies there is instead an excess of
nuclear stellar disks. As the active and inactive
galaxies can be considered two phases of the ``same''  
galaxy, the above findings and dust morphologies suggest
an evolutionary scenario in which external material (gas and
dust) is captured to the nuclear region where it 
settles and ends up feeding
the active nucleus and replenishing the stellar disk -- which
is hidden by the dust in the active galaxies -- with new stars.
This evolutionary scenario is supported by recent
gas kinematics of the inner few hundred parsecs of NGC\,1097,
which shows streaming motions (with velocities $\sim$50\,km\,s$^{-1}$) 
towards the nucleus along spiral arms. The implied large scale
mass accretion rate is much larger than the one derived
in previous studies for the nuclear accretion disk, 
but is just enough to accumulate
one million solar masses over a few million years in the
nuclear region, thus consistent with the recent
finding of a young circumnuclear starburst of one million
solar masses within 9 parsecs from the nucleus in this
galaxy.

\keywords{Galaxies: active -- galaxies: nuclei -- galaxies: kinematics and dynamics --
galaxies: individual (NGC\,1097) -- galaxies: ISM}
\end{abstract}

\firstsection 
\section{Introduction}
The relation between the morphologies of host galaxies and the presence of
nuclear activity has been investigated for many years. One of the first
studies arguing for a difference between active and inactive galaxy hosts
is the one of \cite{simkin80}. These authors have
claimed that Seyfert galaxies hosts are more distorted than inactive galaxies,
showing an excess of bars, rings and tails. More recent studies did not confirm
the excess of bars in Seyferts (e.g. \cite{mulchaey97}) 
while others have found a small excess (e.g. \cite{knapen00}).

A number of studies using Hubble Space Telescope images of nearby galaxies
have revealed a trend for active galaxies always showing a lot of dust structure
in the nuclear region. \cite{vandokkum95} have found that  radio-loud 
early-type galaxies have more dust than radio-quiet ones.
\cite{pogge02} and later \cite{martini03} found that  
Seyfert galaxies almost always present dusty filaments and spirals in the nuclear region,
while \cite{xilouris02} found that, among early Hubble types, 
active galaxies  present more dust structure than inactive galaxies.
Recently, \cite{lauer05} have also 
argued that dust in early-type galaxies is correlated 
with nuclear activity.

The goal of the present paper is to discuss the results of a recent study 
(\cite{lopes07}) which provides a robust analysis 
of the relation between nuclear dust structures
and activity in galaxies on the basis of optical HST
images. The particular case of NGC\,1097
is then discussed, for which we have obtained, in adition,
kinematics of the gas associated with the nuclear dust structures \cite{fathi06}
which reveal streaming motions towards the nucleus.
We then use results of previous studies (\cite{sb03}, \cite{nemmen06})
to relate the mass accretion rate reaching
the accretion disk to the larger scale mass flow rate 
derived from the observed streaming motions.

\section{Correlation between circumnuclear dust and nuclear activity}

We (\cite{lopes07}) have recently selected a sample 
comprising all active galaxies from the Palomar survey \cite{ho95} 
which have optical images in the
HST archive, as well as a pair-matched sample of 
inactive galaxies, and constructed
``structure maps", a technique proposed by \cite{martini99} to enhance 
both absorption and emission structures in the images. 
The total sample comprises 34 matched pairs of early-type
galaxies (T$\le$0) and 31 pairs of late-type galaxies (T$>$0).

The results for the 34 early-type pairs of our sample
are illustrated in Fig. 1 for a subsample of 
10 pairs: while all (100\%) active galaxies present
some dust structure, only 26\% (9 of 34) of the inactive galaxies
possess some dust. Another difference is the
finding that at least 13 of the 34 (38\%) early-type inactive galaxies
present bright {\it stellar disks}, while only
one of the active galaxies show such disks. 
Most dust structures and stellar disks are seen within a few hundred
parsecs from the nucleus.

\begin{figure}
\includegraphics[height=6.2in,width=2.6in]{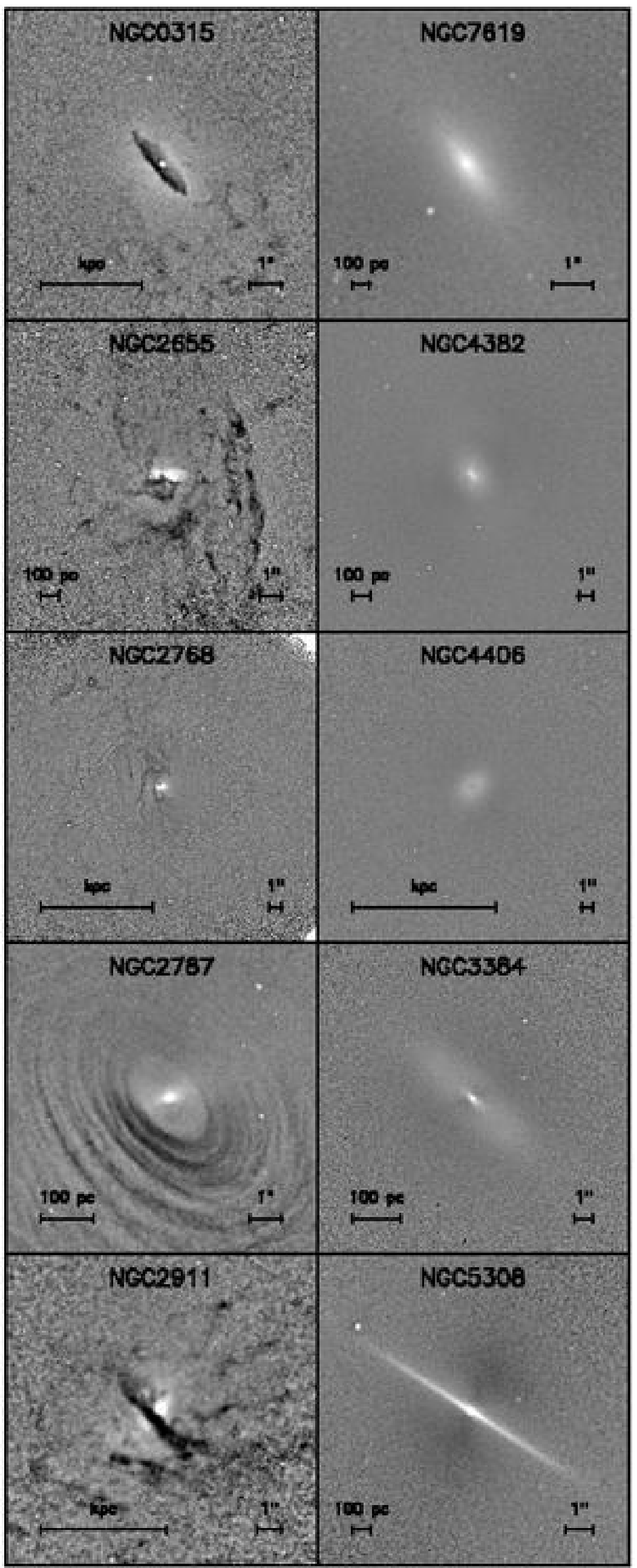}
\includegraphics[height=6.2in,width=2.6in]{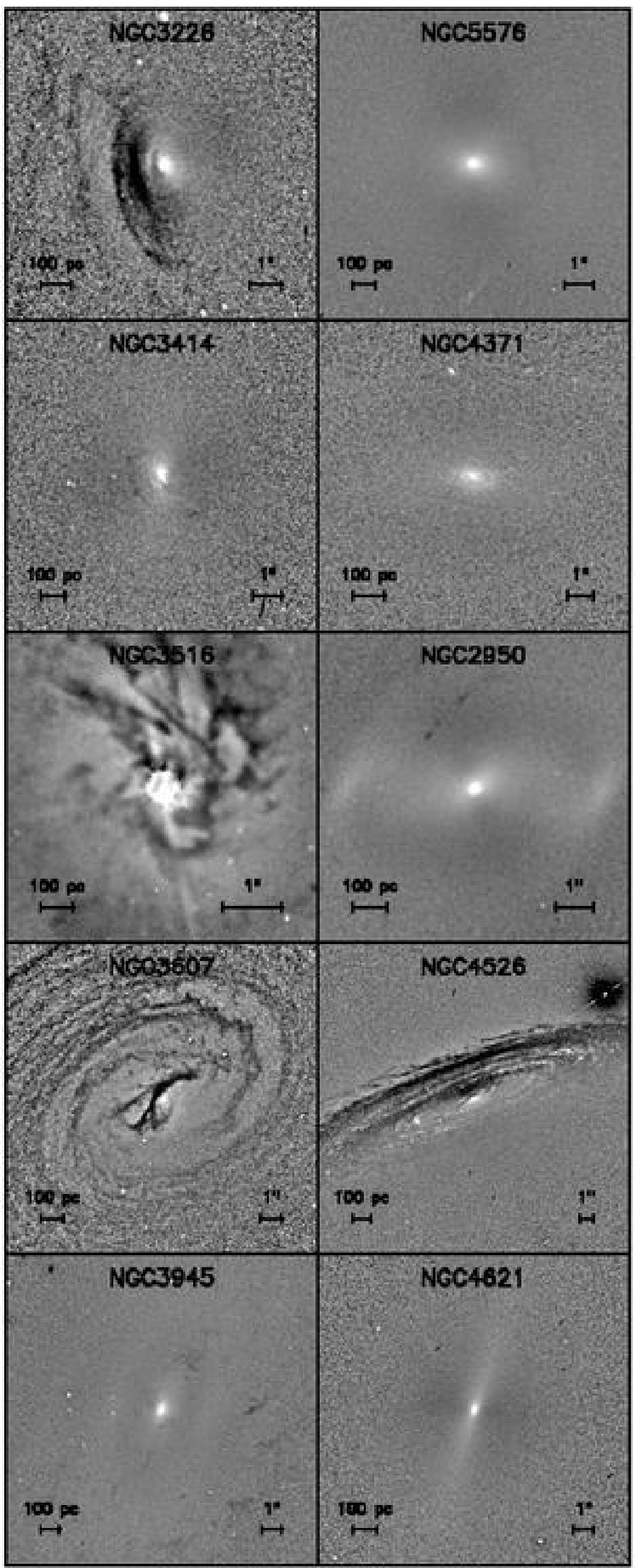}
  \caption{Structure maps for 10 matched pairs of early-type galaxies. 
Each image covers 5\% of the galaxy diameter D25. North is up and East to the left. In each pair the
active galaxy is shown to the left and the inactive matched pair to the right. From \cite{lopes07}.}
\end{figure}
 
A different result was obtained for the late-type pairs:
all galaxies show circumnuclear dust regardless the
presence or absence of nuclear activity, and stellar
disks were found only in a couple of galaxies. 

The above findings imply a strong correlation between
circumnuclear dust and nuclear activity, 
indicating that the dust is connected
to the material which is currently feeding the active
nucleus, and is probably tracing this
material on its way to the nucleus. 
The morphologies of the dust structures range from
chaotic filaments to regular dusty spirals and disks,
suggestive of a ``setling scenario'' for the dust, as proposed by \cite{lauer05}.
Our finding of nuclear stellar disks 
in inactive galaxies suggests that there is one further
step in this evolutionary scenario: the stellar disk,
which shows up in the inactive phase of the galaxy.
The evolutionary scenario can be described as follows:
externally acquired matter is traced by
the chaotic filamentary structure which gradually settles
into more regular nuclear spirals and disks.
Stars then form in the dusty disks, and when the 
gas and dust are fed to the black hole, nuclear activity
ceases and and the stellar disks are unveiled \cite{ferrarese06}.
As stellar disks should be longer lived than the
gasesous dusty disks, the stellar disks are probably
present also in the active phase, but are obscured
by dust. In the evolutionary scenario proposed
above, the episodic accretion of matter then replenishes
the disk, which grows together with the mass
of the nuclear black hole.

On the theoretical side, support for the evolutionary
scenario includes the work of  \cite{maciejewski04} 
who demonstrated that, if a central SMBH  is present, 
nuclear disks of gas and dust can develop spiral shocks 
and  generate gas inflow compatible to the accretion rates observed in 
local active galaxies. On the observational side, kinematic evidence for 
inflow along nuclear dusty spiral arms has been found so far in
one case: NGC\,1097 \cite{fathi06}.

\section{Streaming motions along nuclear spirals in NGC\,1097}

The observations of NGC\,1097 were obtained with
the Integral-Field Unit of the Gemini Multi-Object Spectrograph
and allowed the maping of the gas kinematics in the inner few hundred parsecs
\cite{fathi06}. After subtracting a circular velocity model, 
streaming motions along spiral arms with inward
velocities of up to 50\,km\,s$^{-1}$ were found.
Another relevant finding on this galaxy is the young
obscured starburst recently discovered \cite{sb05} very close to the nucleus
(within $\sim$9\,pc), in agreement with the suggestion that inflowing
gas and dust gives birth to stars in the nuclear dusty spiral or disk \cite{ferrarese06}.

The velocity observed for the streaming motions along the nuclear spiral allows
an estimate of a few Myr for the gas to flow from a few hundred parsecs to the
nucleus. For an estimated gas density of $\sim$500\,protons\,cm$^{-3}$, and
estimated circular cross-section of 3 spiral arms at 100\,pc from the nucleus
(opening angle 20$^\circ$), we conclude that the mass flow rate along the 
nuclear spiral arms in NGC\,1097 is  $\sim$35 times the one at the accretion disk
of dM/dt=1.1$\times$10$^{24}$\,g\,s$^{-1}$ (\cite{nemmen06}), and allows
the accumulation in the nuclear region of $\sim$10$^6$\,M$_{Sun}$ in a few Myr,
which can provide the necessary matter to give birth to the 
nuclear starburst \cite{sb05}.

\section{Concluding remarks}

We have found a strong correlation between circumnuclear dust and 
nuclear activity, and between nuclear stellar
disks and absence of nuclear activity in early-type galaxies.
As both the active and inactive galaxies can be thought of as
the ``same galaxy'' observed in different phases, our findings
suggest an evolutionary scenario in which the nuclear activity
is triggered by capture of dusty gas to the nuclear region.
The origin of the gas and dust is still not clear, but the
absence of dust in the inactive phase suggests that in cannot
originate from continuous mass loss from stars and is most probably
external. In our proposed evolutionary scenario, once
the gas and dust are captured to the nuclear region
they gradually settle into a nuclear spiral or disk, where 
new stars are born while the excess gas and dust is accreted 
by the nuclear SMBH. Replenishment of a nuclear stellar
disk (observed in the inactive galaxies) 
is the final product after activity ceases. As a result,
both the stellar component of the galaxy and its SMBH at the
nucleus grow after each activity cycle, as implied by the
M$\times\sigma$ relation (e.g. \cite{tremaine02}).

The evolutionary scenario is supported by recent kinematic 
observations of the nuclear spiral in NGC\,1097, which 
reveal streaming motions along the spiral arms, providing
an accretion flow rate which is enough to both form 
the starburst recent observed surrounding the nucleus \cite{sb05} 
in a few Myr and feed the nuclear SMBH. As
similar dusty structures
are observed in most active galaxies (\cite{lopes07}), streaming motions
along nuclear spirals may be the main mechanism for
black hole feeding in active galaxies.

\newcommand{\myit}[1]{\rm{#1}}

\bigskip

\discuss{Mitchell Begelman}{Are you able to distinguish flow along a bar
from flow along a spiral? If so, do you see evidence of bars?}

\discuss{Thaisa Storchi-Bergmann}{We would be able to distinguish a bar
from a spiral if there was a bar. We do not see gaseous/dusty bars. 
Bars have been seen in continuum images, but we do not see them in our
structure maps.}

\discuss{Yiping Wang}{Have you done any statistics of your sample to see
a relationship between the star-formation rate and the accretion rate
for the inflow? It seems that star-formation in the nuclear region
and the accretion are linked. We propose a model aimed to explain
the black hole--bulge mass relation in which this relation is important.}

\discuss{Thaisa Storchi-Bergmann}{The 
accretion rate is $\sim0.6M_{\odot}$ per year in NGC\,1097 (the only case
so far in which we were able to calculate the large scale inflow rate), so we can
accumulate enough mass in one million years to form the young star cluster
observed at the nucleus and 
produce necessary star-formation and feeding. We plan more work on our
sample -- this is very important, but we do not have enough data right
now to perform such analysis.}

\end{document}